\documentclass[12pt]{article}
\usepackage{a4wide}
\usepackage{amssymb}
\begin{document}
{\renewcommand{\thefootnote}{\fnsymbol{footnote}}
\begin{center}
{\LARGE  The BKL scenario, infrared renormalization,\\ and quantum cosmology}\\
\vspace{1.5em}
Martin Bojowald\footnote{e-mail address: {\tt bojowald@gravity.psu.edu}}
\\
\vspace{0.5em}
Institute for Gravitation and the Cosmos,\\
The Pennsylvania State
University,\\
104 Davey Lab, University Park, PA 16802, USA\\
\vspace{1.5em}
\end{center}
}

\setcounter{footnote}{0}

\begin{abstract}
  A discussion of inhomogeneity is indispensable to understand quantum
  cosmology, even if one uses the dynamics of homogeneous geometries as a
  first approximation. While a full quantization of inhomogeneous gravity is
  not available, a broad framework of effective field theory provides
  important ingredients for quantum cosmology. Such a setting also allows one
  to take into account lessons from the Belinski--Khalatnikov--Lifshitz (BKL)
  scenario. Based on several new ingredients, this article presents conditions
  on various parameters and mathematical constructions that appear in
  minisuperspace models. Examples from different approaches demonstrate their
  restrictive nature.
\end{abstract}

\section{Introduction}

Quantum gravity is a quantum theory of many interacting degrees of
freedom. Such a theory, in general, requires approximations and assumptions in
order to derive reliable predictions of physical phenomena. Given the
complexity of such a theory, it is hard to find good candidates without
observational assistance. Condensed-matter physics, for instance, provides a
wealth of examples in which this program has been followed through.

In quantum gravity, by contrast, no clear observations will be available for
the foreseeable future.  One could draw the lesson that one should postpone
any quantum-gravity phenomenology until observations can indicate a good
starting point for such an analysis. However, we also need phenomenology to
suggest promising experiments. The main conundrum of quantum gravity is
therefore a chicken-and-egg problem --- what should come first, good
phenomenology that can suggest experiments, or observations that indicate how
to do reliable phenomenology?  Effective field theory can provide a solution:
By parameterizing a large class of potential outcomes, promising effects can
be highlighted.

One realm in which important quantum-gravity effects are expected is
cosmology. Quantum cosmology has traditionally been performed by various
mathematical studies of simple (toy, minisuperspace) models and different
kinds of perturbations around them, but no systematic effective field theory
is available. The purpose of this paper is to point out several ingredients of
quantum cosmology suggested by effective arguments, and to show that some
existing models are at odds with these properties. The main contributions to
this program, which in individual form are not new but appear here in a novel
and fruitful combination, are: (i) A minisuperspace approximation (as opposed
to truncation) \cite{MiniSup}, (ii) the Belinskii--Khalatnikov--Lifshitz (BKL)
scenario \cite{BKL}, and (iii) the infrared behavior of gravity
\cite{InfraredGrav}.

The main ingredient missing in existing models of quantum cosmology is
infrared renormalization, introduced here in the context of cosmological
models. Three approaches will be analyzed in this new picture. While
the effective framework modifies the interpretations of all these examples,
one of them is seen to require a major revision: Loop quantum cosmology, in
its commonly practiced form, does not obey the conditions extracted from
effective field theory.

\section{Minisuperspace approximation}

We introduce the first major deviation from standard quantum cosmology by way
of a brief review of the results of \cite{MiniSup}. As a well-controlled
setting, we consider scalar field theory on Minkowski space-time with
a potential $W(\phi)$ and Lagrangian
\begin{equation} \label{L}
  L=\int{\rm d}^3x
  \left(\frac{1}{2}\dot{\phi}^2-\frac{1}{2}|\nabla\phi|^2 -W(\phi)\right)\,.
\end{equation}
A minisuperspace truncation of this theory is obtained by assuming that $\phi$
is spatially constant, and then integrating over some fixed spatial region
with finite volume $V_0=\int{\rm d}^3x$. The resulting minisuperspace
Lagrangian is
\begin{equation}
 L_{\rm mini}= V_0 \left(\frac{1}{2}\dot{\phi}^2-W(\phi)\right)\,.
\end{equation}
It implies the minisuperspace momentum 
\begin{equation} \label{Mom}
 p=V_0\dot{\phi}\,.
\end{equation}
The minisuperspace Hamiltonian
\begin{equation}\label{H}
 H_{\rm mini}= \frac{1}{2}\frac{p^2}{V_0}+V_0W(\phi)\,.
\end{equation}
is straightforwardly quantized to
\begin{equation} 
 \hat{H}_{\rm mini} = \frac{1}{2}\frac{\hat{p}^2}{V_0}+
   V_0W(\hat{\phi})  \,.
\end{equation}

\subsection{Quantum theory of regions}

Without doing a detailed analysis, it can easily be seen that quantum
corrections, unlike classical effects, depend on the averaging volume
$V_0$. For instance, for a quadratic potential, changing $V_0$ would have the
same effects on quantum corrections as changing the mass has for the standard
harmonic oscillator. This minisuperspace result is conceptually related to
physical effects in quantum-field theory, such as the Casimir force between
two plates which depends on the size of an enclosed region.  Our first lesson,
which will play an important role in the effective field theory to be outlined
in what follows, is therefore that quantum cosmology is a quantum theory of
regions (or patches in the description given in \cite{ROPP}). It is not a
quantum theory of the metric or scale factor, as it is often presented. In
particular, quantum effects in homogeneous models depend on the apparently
arbitrary size $V_0$ of an averaging region.

A variety of methods can be used to derive the effective potential
\begin{equation} \label{Wmini}
 W^{\rm mini}_{\rm eff}(\phi)= W(\phi)+\frac{1}{2V_0} \hbar \sqrt{W''(\phi)}
\end{equation}
in the minisuperspace model to first order in $\hbar$, clearly showing the
$V_0$-dependence of quantum corrections. For instance, the leading term of the
low-energy effective action applied to quantum mechanics as a
$0+1$-dimensional quantum-field theory is of this form \cite{EffAcQM}, and an
independent derivation can be done using canonical effective methods
\cite{EffAc,Karpacz}.

The canonical derivation illustrates the role of quantum fluctuations: If we
take the expectation value of the Hamiltonian $\hat{H}$ in a semiclassical
state with fluctuations $(\Delta\phi)^2\sim\hbar$ and $(\Delta p)^2\sim\hbar$,
to first order in $\hbar$ we can write
\begin{eqnarray}
 \langle\hat{H}\rangle &=& \frac{\langle\hat{p}^2\rangle}{2V_0}+ V_0 \langle
 W(\hat{\phi})\rangle\nonumber\\
  &=& \frac{\langle\hat{p}\rangle^2}{2V_0}+ \frac{(\Delta
   p)^2}{2V_0}+ V_0 W(\langle\hat{\phi})\rangle+ \frac{1}{2}V_0
 W''(\langle\hat{\phi}\rangle) (\Delta\phi)^2+\cdots \label{HExp}
\end{eqnarray}
Heisenberg's equations of motion can be used to derive the following time
derivatives of fluctuations, coupled to the covariance $\Delta(\phi
p)=\frac{1}{2}\langle\hat{\phi}\hat{p}+\hat{p}\hat{\phi}\rangle-
\langle\hat{\phi}\rangle\langle\hat{p}\rangle$:
\begin{eqnarray}
 \frac{{\rm d}(\Delta\phi)^2}{{\rm d}t} &=& \frac{2}{V_0}\Delta(\phi
 p)+\cdots\label{dDeltaq} \\
 \frac{{\rm d}\Delta(\phi p)}{{\rm d}t} &=& \frac{1}{V_0}(\Delta p)^2-
 V_0W''(\langle\hat{\phi}\rangle) (\Delta \phi)^2+\cdots\label{dDeltaqp}\\
 \frac{{\rm d}(\Delta p)^2}{{\rm d}t} &=& -2V_0W''(\langle\hat{\phi}\rangle)
 \Delta(\phi p)+\cdots \label{dDeltap}
\end{eqnarray}
again to first order in $\hbar$.  For an expansion around the stationary
ground state, the moments are (almost) constant in time. For simplicity, we
set the time derivatives exactly equal to zero. Small variations in time can
be included by a systematic adiabatic expansion
\cite{EffAc,Karpacz,HigherTime}. Therefore, 
\begin{equation} \label{Deltaqp}
 \Delta(\phi p)=0
\end{equation}
from (\ref{dDeltaq}) or (\ref{dDeltap}) and
\begin{equation}\label{pphi}
 (\Delta p)^2= V_0^2 W''(\langle\hat{\phi}\rangle) (\Delta\phi)^2
\end{equation}
from (\ref{dDeltaqp}).  We then minimize the contribution
\begin{equation}
 H_{\Delta}= \frac{(\Delta p)^2}{2V_0}+\frac{1}{2}
 V_0W''(\langle\hat{\phi}\rangle) (\Delta\phi)^2
\end{equation}
of fluctuations to the Hamiltonian (\ref{HExp}), again to be close to the
ground state, respecting the uncertainty relation
\begin{equation} \label{Uncert}
 (\Delta\phi)^2(\Delta p)^2-\Delta(\phi p)^2\geq \frac{\hbar^2}{4}\,.
\end{equation}
Since $H_{\Delta}$ is linear in fluctuations, the minimum is realized at the
boundary implied by the inequality (\ref{Uncert}), or for fluctuations
saturating the uncertainty relation. Combined with (\ref{pphi}), we obtain
\begin{equation}
 (\Delta p)^2= \frac{\hbar^2}{4(\Delta\phi)^2}=
 V_0^2W''(\langle\hat{\phi}\rangle) (\Delta\phi)^2
\end{equation}
or 
\begin{equation}
 (\Delta \phi)^2=\frac{\hbar}{2V_0
 \sqrt{W''(\langle\hat{\phi}\rangle)}}\quad,\quad (\Delta p)^2=
\frac{1}{2}V_0\hbar \sqrt{W''(\langle\hat{\phi}\rangle)}\,.
\end{equation}
With these values, $H_{\Delta}/V_0$ equals the correction term in
(\ref{Wmini}). As shown by this derivation, the $V_0$-dependence of quantum
corrections is determined by two universal properties: The symplectic
structure or canonical relationships used to derive Heisenberg's equations of
motion (\ref{dDeltaq})--(\ref{dDeltap}), together with the uncertainty
relation. In the next section, we will confirm the same qualitative behavior
in quantum cosmology. But first we have to find a valid interpretation of
$V_0$ within effective field theory.

\subsection{Infrared contributions}

The full theory (\ref{L}), from which we derived the minisuperspace model,
also has an effective potential: the Coleman--Weinberg potential
\cite{ColemanWeinberg}
\begin{equation}
 W_{\rm eff}(\phi) = W(\phi)+ \frac{1}{2}\hbar \int
 \frac{{\rm d}^4k}{(2\pi)^4} \log \left(1+
   \frac{W''(\phi)}{||{\bf k}||^2}\right)\,.
\end{equation}
It looks very different from the minisuperspace effective potential, but, as
noticed in \cite{CW}, performing the $k^0$-integration in closed form reveals
their similarity: The Coleman--Weinberg potential then equals
\begin{equation} \label{W}
 W_{\rm eff}(\phi) = W(\phi)+ \frac{1}{2}\hbar \int
 \frac{{\rm d}^3k}{(2\pi)^3}  \left(\sqrt{|\vec{k}|^2+W''(\phi)}-
   |\vec{k}|\right) \,.
\end{equation} 

The minisuperspace potential is related to the field-theory effective
potential through the infrared contribution
\begin{equation} \label{WIR}
 W_{\rm IR}(\phi)=\frac{1}{2}\hbar \int_{|\vec{k}|\leq k_{\rm max}}
 \frac{{\rm d}^3k}{(2\pi)^3}  \left(\sqrt{|\vec{k}|^2+W''(\phi)}-
   |\vec{k}|\right)
\end{equation}
of the latter: If we use quantum-field theory to describe the modes with wave
length greater than $1/V_0^{1/3}$, which remain inhomogeneous after averaging
over a region with volume $V_0$, they result in an infrared contribution with
$k_{\rm max}= 2\pi/V_0^{1/3}$. For large $V_0$ and therefore small $k_{\rm
  max}$, we can replace the integrand in (\ref{WIR}) with the integrand at
$\vec{k}=0$ times the $k$-integration volume:
\begin{equation} \label{W2}
 W_{\rm IR}(\phi) \approx W(\phi)+ \frac{\hbar}{12
  \pi^2} k_{\rm max}^3 \sqrt{W''(\phi)}= W(\phi)+\frac{2\pi}{3V_0}\hbar
\sqrt{W''(\phi)} 
\end{equation}
in agreement with $W_{\rm eff}^{\rm mini}$ up to a numerical factor. The
different numerical factor can be related to the separation of modes, which is
more clear in models with a discrete spectrum of $\vec{k}$. Such models,
studied in more detail in \cite{MiniSup}, can lead to complete agreement
between minisuperspace potentials and infrared contributions of field-theory
potentials. Here, we are mainly interested in seeing the common
$V_0$-dependence.

\section{Infrared behavior}

A more complicated infrared behavior is obtained in theories with massless
excitations, such as gravity. In the scalar model, we have implicitly assumed
that $W''(\phi)$ is sufficiently large, such that
$\sqrt{|\vec{k}|^2+W''(\phi)}$ can be replaced in (\ref{W2}) by its Taylor
expansion with respect to $|\vec{k}|$ in the entire integration region up to
$k_{\rm max}$. If $W''(\phi)\ll |\vec{k}|^2$, however,
\begin{equation} \label{ExpIR}
 \sqrt{|\vec{k}|^2+W''(\phi)} \approx |\vec{k}|+\frac{1}{2}
\frac{W''(\phi)}{|\vec{k}|}
\end{equation}
is different from the previous expansion. Evaluated at $k_{\rm max}$ (since
(\ref{ExpIR}) is singular at $|\vec{k}|=1$) and multiplied with the
$k$-integration volume in (\ref{WIR}), we have
\begin{equation}
 W_{\rm IR}(\phi)\approx W(\phi)+ \frac{\hbar}{12\pi^2}\left( k_{\rm max}^4+
   \frac{1}{2} W''(\phi) k_{\rm max}^2\right) \,.
\end{equation}
In particular,
\begin{equation}
 W_{\rm IR}(\phi) \approx W(\phi)+ \frac{4}{3}\pi^2\hbar V_0^{-4/3}+
 \frac{1}{6}\hbar W''(\phi)V_0^{-2/3}
\end{equation}
contains terms with a different powers of $V_0$, compared with (\ref{W2}),
which can be important for large averaging volumes. If (\ref{ExpIR}) is pushed
to higher orders in $1/|\vec{k}|$, positive powers of $V_0$ even appear in
$W_{\rm IR}(\phi)$. However, the entire infrared expansion of a massless
theory requires a more careful derivation.

The problem of infrared contributions to massless theories is that both
$W''(\phi)$ and $|\vec{k}|$ are small, and no obvious expansion can be
done. Although the naive expansion (\ref{ExpIR}) could then be misleading, it
turns out that it does indicate the correct qualitative properties in an
application to gravity: The detailed analysis given in \cite{InfraredGrav}
shows that the gravitational effective potential $W$ has an infrared fixed
point where it has a small-$|\vec{k}|$ expansion of the form
\begin{equation} \label{Wcc}
 W=\frac{c_1}{16\pi G} |\vec{k}|^2- \pi G c_2|\vec{k}|^6
\end{equation}
(slightly adapted to our notation) with $k$-independent $c_1$ and $c_2$, which
depend on background values through a parameterization of the infrared
flow. The effective potential contributes a term 
\begin{equation} \label{V0W}
 V_0W=\frac{8\pi^3}{|\vec{k}|^3} W= \frac{\pi^2}{2G}\frac{c_1}{|\vec{k}|}-
 8\pi^4G c_2|\vec{k}|^3
\end{equation}
to the Hamiltonian constraint, which receives a term proportional to
$|\vec{k}|^3\propto 1/V_0$ from (\ref{Wcc}) just as in (\ref{W2}), but also a
term proportional to an inverse power $|\vec{k}|^{-1}$.  In particular, the
infrared contribution to the Hamiltonian constraint is not a Taylor series in
$|\vec{k}|$, in agreement with (\ref{ExpIR}).

\subsection{Canonical quantum cosmology}

We can see the same feature independently in a generic analysis of
quantum-cosmological models. Quantum corrections with inverse powers of
$|\vec{k}|$ should then appear as positive powers of $V_0$, unlike what is
seen in a minisuperspace effective potential such as (\ref{Wmini}). At the
same time, we can make a connection with the main lesson from the canonical
derivation of the effective potential (\ref{Wmini}): the interplay of
canonical relationships with the uncertainty relation.

The relevant canonical relationships are determined by the ADM formulation of
general relativity \cite{ADM}: The spatial metric $h_{ab}$ has momenta given
by 
\begin{equation} \label{pab}
 p^{ab}= \frac{\sqrt{\det h}}{16\pi G} \left(K^{ab}- K^c{}_c h^{ab}\right)
\end{equation}
in terms of extrinsic curvature 
\begin{equation}
 K_{ab}= \frac{1}{2N}\left(\dot{h}_{ab}- D_aN_b-D_bN_a\right)\,.
\end{equation}
For isotropic cosmological models, the shift vector $N^a=0$ and its spatial
covariant derivatives $D_aN_b$ are zero, and we can assume $N=1$ for proper
time. The spatial part $h_{ab}=a^2\delta_{ab}$ of the
Friedmann--Robertson--Walker metric then implies that (\ref{pab}) simplifies
to
\begin{equation}
 p^{ab}= -\frac{\dot{a}}{8\pi G} \delta^{ab}\,.
\end{equation}
The symplectic potential $\int{\rm d}x^3 p^{ab}\delta h_{ab}$ in a
Lagrangian, integrated over the averaging volume of an isotropic model, is
therefore reduced to
\begin{equation}
 \int{\rm d}x^3 p^{ab}\delta h_{ab}= -\frac{3V_0}{8\pi G}\dot{a}\delta a^2
\end{equation}
from which we read off that the momentum canonically conjugate to
$V_0^{2/3}a^2$ is equal to $-(3/8\pi G)V_0^{1/3}\dot{a}$. Here, we have
distributed $V_0$ such that any $a$ is accompanied by a factor of $V_0^{1/3}$,
ensuring that the canonical variables are invariant under rescaling
spatial coordinates.

For quantum cosmology, we use a general parameterization of basic canonical
variables. In the presence of ambiguities, we need to describe quantization
choices related to the representation of the scale factor $a$ and its momentum
as operators. We begin with the canonical pair just derived, and apply a
1-parameter family of canonical transformations such that the momentum remains
linear in $\dot{a}$ but $a$ appears in different power laws:
\begin{equation} \label{QP}
 Q=\frac{3}{8\pi G} \frac{(V_0^{1/3}a)^{2(1-x)}}{1-x}\quad,\quad P=
-(V_0^{1/3}a)^{2x}V_0^{1/3}\dot{a}
\end{equation}
with a real parameter $x$, such that $\{Q,P\}=1$ for any $x$.  The scale
factor as basic variable is obtained from (\ref{QP}) for $x=1/2$, while the
volume corresponds to $x=-1/2$. The only common choice not strictly included
in this parameterization is the logarithmic variable $\log a$, canonically
conjugate, up to a multiplicative constant, to $a^2\dot{a}$. Formally, this
choice can be obtained from (\ref{QP}) in the limit $x\to 1$.

We can already see that these canonical relationships together with the
uncertainty relation restrict possible $V_0$-dependences: We have $QP\propto
V_0$ for any $x$, while $\Delta Q\Delta P\geq \hbar/2$ is bounded from below
by a $V_0$-independent constant. The ``semiclassicality'' parameters
$Q^{-1}\Delta Q$ and $P^{-1}\Delta P$ must therefore depend on $V_0$, and so
do quantum corrections to an effective Hamiltonian.

In a canonical effective theory \cite{EffAc,Karpacz}, the basic variables $Q$
and $P$ correspond to expectation values of basic operators, while
fluctuations are independent quantum variables characterizing a state. These
variables, as well as higher moments, are responsible for quantum corrections
as in (\ref{Wmini}). Their precise form, such as a dependence on $a$ for
solutions of the theory, would require a detailed dynamical analysis,
replacing information provided in standard systems by the no longer existing
ground state. For our purposes it is sufficient to continue with parameterized
equations. In particular, we assume that
\begin{equation}
 (\Delta Q)^2\propto (V_0^{1/3}a)^{4y}
\end{equation}
with a new, generically non-zero constant $y$. We need not make assumptions
about $\Delta P$ because it is related to $\Delta Q$ by the uncertainty
relation $(\Delta Q)^2(\Delta P)^2\geq \hbar^2/4$. If the state is nearly
semiclassical (in a broad sense, that is, not necessarily Gaussian), it is
close to saturating the uncertainty relation. For $y\not=0$, $\Delta Q$
decreases in forward or backward evolution of an expanding universe, and
we will quickly violate the uncertainty relation unless $\Delta P$ changes
suitably. The saturation limit is respected if
\begin{equation}
 (\Delta P)^2\propto (V_0^{1/3}a)^{-4y}\,,
\end{equation}
such that $(\Delta Q)^2(\Delta P)^2$ is constant.  Importantly, the
uncertainty relation, imposing a $V_0$-independent lower bound, $\hbar^2/4$,
implies that the product $\Delta Q \Delta P$ has a dependence on $V_0$
different from the product $QP\propto V_0a^2\dot{a}$. This general fact is
responsible for the $V_0$-dependence of quantum corrections.

Combining the preceding equations, we can derive the parameterized scaling of
quantum corrections in the Hamiltonian constraint of a quantum cosmological
model. The classical constraint $H(Q,P)$, like (\ref{H}), is such that every
term scales like $V_0$. For dimensional reasons, the leading quantum
corrections linear in $(\Delta Q)^2$ and $(\Delta P)^2$ are of the form
$(\Delta Q)^2\partial^2 H/\partial Q^2$ and $(\Delta P)^2\partial^2 H/\partial
P^2$. Since $H$ scales like $V_0$, the scaling behavior of the corrections can
be read off from $V_0 (\Delta Q)^2/Q^2$ and $V_0(\Delta P)^2/P^2$. In our
parameterization,
\begin{equation} \label{QPrel}
 V_0\frac{(\Delta Q)^2}{Q^2}\propto V_0^{4(x+y-1/4)/3}
\propto
 k_{\rm max}^{1-4(x+y)}\quad,\quad V_0\frac{(\Delta P)^2}{P^2}
\propto V_0^{-4(x+y-1/4)/3} \propto k_{\rm max}^{4(x+y)-1}\,.
\end{equation}
As a consequence of how we have chosen our parameterization, together with the
uncertainty relation, these quantities depend on a single parameter, $x+y$.
It is now easy to see that, unless $x+y=1/4$, the leading quantum corrections
in the Hamiltonian constraint of quantum cosmology contain an inverse power of
$k_{\rm max}$, from $V_0(\Delta Q)^2/Q^2$ if $x+y>1/4$ and from $V_0(\Delta
P)^2/P^2$ if $x+y<1/4$.  This statement agrees with (\ref{V0W}) derived from
the detailed analysis of \cite{InfraredGrav}. The $V_0$-behavior of quantum
corrections depends on quantization choices (through $x$) and quantum dynamics
via the behavior of fluctuations (through $y$).

\subsection{Infrared renormalization}

Having established a close relationship between quantum corrections in
minisuperspace models and infrared contributions to quantum-field theories, we
return to the minisuperspace approximation. Quantum corrections in
minisuperspace models, implicitly, give an approximate description of
interactions of those modes of the full quantum-field theory that have not
been averaged out, and therefore have wavelengths greater than the averaging
volume $V_0$ of the minisuperspace model. This conclusion is based on the two
main steps in our derivation of this relationship: To obtain (\ref{W2}) from
(\ref{W}), we (i) include only modes with $|\vec{k}|\leq k_{\rm max}=
2\pi/V_0^{1/3}$, and (ii) replace the remaining mode integral by its integrand
evaluated at small $|\vec{k}|$, multiplied by a small $k$-volume. The
minisuperspace approximation can therefore be expected to be reliable in
regimes with
\begin{enumerate}
\item[(i)] significant inhomogeneity only on scales greater than $V_0$,
  provided that
\item[(ii)] $V_0$ is large.
\end{enumerate}  
Both conditions are fulfilled in late-universe cosmology, averaging over a
Hubble region, but in quantum cosmology we are usually more ambitious and aim
to apply quantum theory to the early universe, or even to understand the
big-bang singularity.

As we approach a spacelike singularity, for instance in backward evolution
from our present nearly homogeneous state, distance scales of structure, and
therefore any region which is approximately homogeneous, shrink with a
decreasing $a$. However, inhomogeneity grows even within a comoving
region. Existing structure is not only brought to smaller distances by a
shrinking scale factor, it is also enhanced by gravitational collapse that
forms new structure. If we try to describe the classical dynamics underlying a
minisuperspace model, therefore, inhomogeneity grows within any region of
constant $V_0$.

In order to maintain the minisuperspace description, we should then gradually
shrink $V_0$ as we evolve toward a spacelike singularity. Since we have
associated the averaging volume with an infrared scale, adjusting $V_0$
amounts to infrared renormalization. It is important to understand how this
process affects the approach to a singularity, as seen in a homogeneous
model. At this stage, the BKL scenario enters the picture.  When we get close
to a spacelike singularity, the BKL scenario \cite{BKL} sets in as an
asymptotic statement. It tells us that we can assume a homogeneous geometry
right up to the spacelike singularity, a conclusion which is often cited as a
justification of minisuperspace truncations; see for instance
\cite{BounceKasner} as a recent example. However, as an asymptotic statement,
the BKL scenario does not place any lower limit on $V_0$, not even the Planck
volume.  Since $V_0$ has to decrease on approach to the singularity in order
to maintain the minisuperspace assumption, generically we should therefore
describe a geometry close to a spacelike singularity using small $V_0$: We
obtain a local homogeneous geometry, such as Bianchi IX, that describes how
the metric changes at a given point, but not a full Bianchi IX model which
includes the topological space on which it is formulated. There is an
important difference between these two applications of homogeneous solutions
because quantum cosmology, as we have learned, is a quantum theory of
regions. It therefore matters whether we can assume a dynamical behavior only
locally or for a global space.

In addition, the shrinking $V_0$ implies that the minisuperspace approximation
becomes less and less reliable near a spacelike singularity: Condition (ii) at
the beginning of this subsection is then violated. Even though the BKL scenario
allows us to use the classical dynamics of homogeneous models to understand
space-time near a spacelike singularity, it is a poor justification of
minisuperspace models in quantum cosmology. Using a minisuperspace model to
evolve from a nearly homogeneous geometry at late times to a BKL-like geometry
at early times means that we begin with a well-justified, approximate infrared
contribution of the full theory, but then push the infrared scale all the way
into the ultraviolet.

As a technical note, we should expect mixed states in a quantum-mechanical
implementation of reducing the averaging volume, while most studies in quantum
cosmology are based on pure states. Moreover, there is no unitary
transformation that could be used to change $V_0$ in quantum cosmology, since
even a classical change of $V_0$ would not be a canonical transformation; see
the $V_0$-dependence in (\ref{QP}). Therefore, quantum cosmology in a
minisuperspace approximation cannot be based on a single equivalence class of
Hilbert-space representations.  Both features --- the appearance of mixed
states and the impossibility of using a single Hilbert space --- indicate that
effective field theory is required for a proper analysis.

\section{Examples}

Our general discussion can be applied to various approaches to quantum
cosmology, often with an important change in viewpoint.

\subsection{Bohmian quantum cosmology}

As reviewed for instance in \cite{BohmQCRev} Bohmian quantum mechanics
\cite{BohmI,BohmII} applied to a cosmological model with Hamiltonian
\begin{equation}
 H=V_0\left(\frac{1}{2} f^{ab}(q)p_ap_b+U(q)\right)
\end{equation}
implies a quantum potential
\begin{equation}
 W_{\rm Q}=-\frac{V_0}{2\sqrt{f}|\psi|} \frac{\partial}{\partial
   q^a}\left(f^{ab}\sqrt{f}\frac{\partial}{\partial q^b}|\psi|\right)
\end{equation}
derived from the wave function $\psi$.  In this version of quantum cosmology,
a logarithmic basic variable is used, which in terms of $V_0$ behaves
similarly to the scalar example we used to motivate the minisuperspace
approximation. Using the $V_0$-dependence of the momentum (\ref{Mom}) and the
Hamiltonian (\ref{H}), we therefore have
\begin{equation}
 f^{ab}\propto V_0^{-2}\,,
\end{equation}
such that
\begin{equation}
 W_{\rm Q}\propto V_0^{-1} 
\end{equation}
as in (\ref{Wmini}).  Any effects based on the quantum potential in quantum
cosmology are therefore enhanced by infrared renormalization as we evolve to
smaller volumes, closer toward a spacelike singularity.  Bounce models based
on Bohmian quantum cosmology, such as \cite{QCPerfectBohm,NonSingBohmQC}, are
then more secure in the effective picture developed here --- provided the
minisuperspace truncation is reliable.

\subsection{Affine quantization}

In \cite{AffineSing}, affine quantization \cite{AffineQG,AffineQG2} has been
applied to derive an effective Friedmann equation
\begin{equation}
 \frac{\dot{a}^2}{a^2}+\frac{k}{a^2}+ \frac{k_2}{\ell^2a^6}= \frac{8\pi
   G}{3} \rho(a)
\end{equation}
with an effective energy density
\begin{equation}
 \rho(a)=\frac{k_3 \hbar(N+1)}{V_0a^4}
\end{equation}
from harmonic anisotropies, using $\ell = V_0/\ell_{\rm P}^2$. In
\cite{AffineSing}, $V_0$ has been assumed to equal the coordinate volume of a
Bianchi IX space, but we can easily adapt the equations to a running $V_0$
according to infrared renormalization.

The behavior of $\rho(a)$ agrees with our $\hbar \sqrt{W''}/V_0$ in
(\ref{Wmini}). In addition to this effective matter term, there is a repulsive
\begin{equation}
 k_2\ell^{-2}a^{-6}\propto \frac{\hbar^2}{V_0^2}\,,
\end{equation}
which, for $k_2<0$, is able to cause a bounce because it dominates all other
terms for small $a$. This domination is enhanced for small $V_0$ if the model
is combined with infrared renormalization. Also here, the bounce is more
secure in the effective picture.  (However, at present it does not seem clear
whether higher-order $(\hbar/V_0)$-corrections in $\rho(a)$ might compete with
the repulsive term.)

\subsection{Loop quantum cosmology}

An effective Friedmann equation \cite{AmbigConstr}
\begin{equation}\label{Hol}
 \frac{\dot{a}^2}{a^2} = \frac{8\pi G}{3} \rho\left(1-\frac{\rho}{\rho_{\rm
       QG}}\right)
\end{equation}
with
\begin{equation}
 \rho_{\rm QG}(a) = \frac{3}{8\pi G\delta^2 (V_0^{1/3}a)^{2(1+2x)}}
\end{equation}
can, under certain assumptions about the matter ingredients and properties of
a state, be derived from loop quantum cosmology \cite{ROPP}. The parameter
$\delta$, with units of length to the power $-2x$, characterizes the strength
of spatially non-local effects in the theory implied by using
holonomies. Following \cite{APS}, many studies of this and related equations
have been published, but usually assuming a macroscopic value of $V_0$. The
authors of \cite{APS} have argued that late-time homogeneity justifies such a
choice. But as we have seen here, using this postulate throughout long-term
evolution up to high curvature is not compatible with the BKL scenario and an
effective description which, through infrared renormalization, requires that
$V_0$ be adjusted to smaller and smaller values as gravitational collapse
proceeds.

The only reliable information we have about inhomogeneity in cosmology,
relevant for quantum-cosmological models, is late-time near-homogeneity and
the asymptotic statement of the BKL scenario. A bounce somewhere near
Planckian curvature does not fall into either if these two regimes, but
reaching it from well-understood late times certainly requires long evolution
through dense phases with significant gravitational collapse. The assumption
that large $V_0$ can still be used close to Planckian curvature is therefore
very restrictive. There may be cosmological solutions which can be
approximated by a minisuperspace model with constant and large $V_0$ all the
way to the Planck density, but insisting on this assumption amounts to a high
degree of fine-tuning of initial data that describe the late-time geometry.

While it is possible to begin evolving with large $V_0$ at late times, this
parameter must take on smaller and smaller values in order to maintain the
minisuperspace assumption as one approaches high curvature.  For $x>-1/2$, the
correction term $\rho/\rho_{\rm QG}\propto V_0^{2(1+2x)/3}$ decreases for
smaller $V_0$. In contrast to the first two examples, effects that may lead to
a bounce in loop quantum cosmology therefore become weaker as a consequence of
infrared renormalization. 

The borderline case $x=-1/2$ implies corrections in (\ref{Hol}) independent of
$V_0$, but the dynamics remains sensitive to quantum fluctuations which are
not included in (\ref{Hol}): The general effective Friedmann equation deirved
in \cite{BounceSqueezed,QuantumBounce} shows that quantum fluctuations and
correlations contribute to (\ref{Hol}) by modifying the parenthesis
$(1-\rho/\rho_{\rm QG})$ such that
\begin{equation}\label{Holsigma}
 \frac{\dot{a}^2}{a^2} = \frac{8\pi G}{3} \rho\left(1-\frac{\rho}{\rho_{\rm
       QG}}+\sigma\right)
\end{equation}
where 
\begin{equation}
 \sigma= \frac{(\Delta Q)^2-C+\delta^2\hbar^2/4}{(Q+\delta\hbar/2)^2}
\end{equation}
with a correlation parameter $C$. To be specific, we may assume that
$\delta\sim \ell_{\rm P}^{-2x}$ is related to the Planck length $\ell_{\rm
  P}$, as usually done in models of loop quantum cosmology. It is now
important to remember that $Q$ is proportional to a positive power of $V_0$
for values of $x$ usually considered in loop quantum cosmology, in particular
for $x=-1/2$.  For large $V_0$ and a semiclassical state with $C\sim\Delta
Q\sim \delta \hbar$, $\sigma\propto V_0^{-\frac{4}{3}(1-x)}\ll 1$ is
negligible. However, when $V_0$ has reached a small value after infrared
renormalization such that $GQ\ll \delta G\hbar\sim \ell_{\rm P}^{2(1-x)}$, we
have $\sigma\sim 1$ even if the state remains semiclassical. For states at
small $V_0$ that are not semiclassical, as may be expected in a high-curvature
phase, $\sigma$ can be significantly greater than one.  Fluctuations at small
$V_0$ therefore significantly alter the effective Friedmann equation for
densities close to $\rho_{\rm QG}$, where semiclassical large-$V_0$ solutions
would provide a bounce.  In particular, non-bouncing solutions do exist when
small $V_0$ are considered \cite{BouncePert}, even in simple models in which
one can show that all large-$V_0$ solutions bounce. Effective field theory
therefore suggests a significant revision of the conclusions drawn in loop
quantum cosmology following \cite{APS}, based on the assumption that large
$V_0$ can be used throughout the entire evolution.

\section{Conclusions}

A possible effective field theory of quantum cosmology combines the BKL
scenario with effects from quantum field theory and infrared
renormalization. These considerations mainly apply to models in which the
early universe is treated as a transition phase, in particular to bounce
models. Models which treat the early unniverse as an initial stage, such as
the tunneling \cite{tunneling} or the no-boundary proposal \cite{nobound} as
well as recent applications of loop quantum cosmology to such scenarios
\cite{NoBoundLQC,LoopsRescue}, behave differently: At the initial stage, the
scale factor reaches the value $a=0$ such that the entire space, of any size
$V_0$, uniformly collapses to zero size. In these models, $V_0$ is not required
to take on small values.

As an important application to quantum cosmology, our effective description
shows qualitative differences between various approaches that have led to
bounce models.  In particular, it strengthens quantum corrections based on
fluctuations, but also reveals spurious effects, in particular in loop quantum
cosmology where large $V_0$, or large-scale homogeneity within a comoving
volume, is often assumed even for early-universe models.

At the same time, the effective description highlights the main problem of
minisuperspace models: How do we reconcile the necessity of small homogeneous
regions in the asymptotic regime of BKL with the infrared truncation of quantum
field theory implied by a minisuperspace approximation?

\section*{Acknowledgements}

This work was supported in part by NSF grant PHY-1607414.


\end{document}